\documentclass[preprint2]{emulateapj}

\slugcomment{Accepted for publication in ApJS on October 11, 2011}

\begin{document}

\title{Lack of Inflated Radii for {\it Kepler} Giant Planet Candidates Receiving Modest Stellar Irradiation}

\author{Brice-Olivier Demory and Sara Seager
}
\altaffiltext{}{Department of Earth, Atmospheric and Planetary Sciences, Massachusetts Institute of Technology, 77 Massachusetts Ave., Cambridge, MA 02139, USA. demory@mit.edu}

\begin{abstract}

  The most irradiated transiting hot Jupiters are characterized by
  anomalously inflated radii, sometimes exceeding Jupiter's size by more than
  60\%. While different theoretical explanations have been applied,
  none of them provide a universal resolution to this
  observation, despite significant progress in the past years.
  We refine the photometric transit light curve analysis of 115
  {\it Kepler} giant planet candidates based on public Q0-Q2
  photometry. We find that 14\% of them are likely false positives,
  based on their secondary eclipse depth.
  We report on planet radii vs. stellar flux. We find an increase
  in planet radii with increased stellar irradiation for the {\it Kepler} giant
  planet candidates, in good agreement with existing hot Jupiter systems.
  We find that in the case of modest irradiation received
  from the stellar host, giant planets do not have inflated radii, and
  appear to have radii independent of the host star incident flux. This
  finding suggests that the physical mechanisms inflating hot
  Jupiters become ineffective below a given orbit-averaged stellar irradiation level of
  $\sim$2$\times 10^8$ erg s$^{-1}$ cm$^{-2}$. 

\end{abstract}


\section{Introduction}

The onset of giant exoplanet transit science, initiated by the
discovery of HD\,209458\,b \citep{Charbonneau2000, Henry2000},
immediately revealed unexpected anomalous radii for several hot
Jupiters . The so-called inflated radii became common as new transiting hot Jupiters were
discovered. The anomalous giant planet radii
were unexpected because in the mass regime of giant planets,
the mass-radius relationship for giant planets was expected to be
unique \citep{Zapolsky1969}, assuming a given composition. Moreover,
the compensating effects of electron degeneracy and electrostatic
contribution from the classical ions yield a quasi-constant radius
around 1 and 7 $M_{Jup}$ \citep[see, e.g.,][]{Chabrier2009}. Soon after the
discovery of 51\,Peg\,b \citep{MayorQueloz1995}, 
\citet{Guillot1996} correctly pointed out that strongly irradiated giant planets 
do not follow the same mass-radius relationship than the one of isolated objects.

More than 100 known transiting hot Jupiters later, planet radius discrepancies
are still common \citep[e.g.,][]{Baraffe2010,Fortney2010}.
Despite numerous theoretical studies \citep[see,e.g.,][for a
review]{FortneyNettelmann2010}, no universal mechanism seems to
fully account for the observed radii anomalies. 

The motivation of the present work is that most of the proposed
mechanisms to explain inflated radii are expected to become less
effective as the stellar incident flux decreases
\citep[e.g.,][]{Fortney2007}. We therefore study a {\it Kepler}
subsample of 138 giant planets candidates to better understand the
effect of irradiation across a wide range of orbital separation on
giant extrasolar planets radii. We further note that transiting giant planets receiving modest stellar 
irradiation are particularly important for the derivation of their internal 
structure and composition, as the interior energy source is expected 
to affect the planetary radius only nominally \citep{MillerFortney2011}.

The paper is organized as follows. We first describe in Sect.~2 how
the sample of giant planet candidates was chosen. Then we present the
data analysis that refined the system parameters and provided
constraints on planethood of the candidates. The results of this
analysis are then shown in Sect.~3. We finally discuss the behavior
of the {\it Kepler} giant planet candidates in the radii vs.
stellar incident flux plane and estimate the {\it Kepler} false
positive rate for this class of objects in Sect.~4.

\section{Data Analysis}

\subsection{Selection of  giant planet candidates}
This study is based on quarters Q0, Q1 and Q2 {\it Kepler} data that
were publicly released on 2011 February 1st \citep{Borucki2011}. In
total, the datasets encompass 136 days of photometric monitoring
between May and September 2009. 

A list of 1235 {\it Kepler} Objects of Interest (KOI) was released,
unveiling a huge diversity of exoplanet candidates in terms of planetary radii and orbital periods. One of the key
elements of this release is the relatively low occurrence of
Jupiter-sized candidates \citep[see, e.g.,][]{Howard2011}. Out of the 1235 planet candidates, only
about 15\% have radii above 0.6$R_{Jup}$, a result that supports the low frequency of giant planets
found in RV surveys  \citep[e.g.,][]{Howard2010}, albeit for a different stellar population.

Our primary selection criterion is the planetary
radius. \citet{Borucki2011} announced 165 giant planet candidates with
6$R_{\oplus}<R_P<15R_{\oplus}$ and 19 candidates with 15$R_{\oplus}<R_P<22R_{\oplus}$ 
for a total of 184 objects. We
further restricted this sample to keep only ``giant'' planet
 candidates, defined here by 8$R_{\oplus}<R_P<22R_{\oplus}$. This
step yielded 138 candidates. We removed the 14 KOI that have only one
transit in Q0-Q2 and 9 other classified as ``moderate" candidates,
likely false positives, that exhibit centroid motion or
difference of depth between odd and even transits
\citep{Batalha2010}. This selection left us with a sample of 115 giant
planet candidates.

\subsection{Method}

For each of the 115 KOI, we retrieved the Q0-Q2 raw long-cadence
photometry \citep{Jenkins2010a} from
MAST\footnote{http://archive.stsci.edu/kepler/}. These data include
all photometry in the form of individual lightcurves. We used the raw
photometry instead of the Kepler-corrected (PDC) \citep{Jenkins2010b}
photometry so that we could identify systematics on specific
timescales as necessary input to our analysis. Moreover,
by using the raw data we can assess the amplitude of correlated noise
(from instrumental systematics and stellar variability combined) to
derive uncertainties on stellar and planetary parameters.

To better characterize the 115 planet candidates, we performed individual
Markov Chain Monte Carlo (MCMC) analysis for each KOI. The aim of this
analysis was two-fold.  The first goal was to remove false positives by way
of detecting a robust secondary eclipse signal indicative of a stellar
companion instead of a planetary companion. The second goal was to
derive the stellar density from the transit lightcurve in order to
further derive the stellar parameters and planetary radius. This step
also required use of the stellar $T_{eff}$ values drawn from the {\it
  Kepler} Input Catalog \citep[KIC;][]{Brown2011}.

We used the implementation of the MCMC algorithm presented in
\citet{Gillon2009,Gillon2010} in order to derive the stellar and 
planetary parameters. MCMC is a Bayesian inference method
based on stochastic simulations that samples the posterior probability
distributions of adjusted parameters for a given model. Our MCMC
implementation uses the Gibbs sampler and the Metropolis-Hastings
algorithm to estimate the posterior distribution function of all jump
parameters. Our nominal model is based on a star and a transiting
planet on a Keplerian orbit about their center of mass.

Input data provided to each MCMC consisted of the Q0-Q2 {\it Kepler}
photometry and the KIC stellar $T_{eff}$ value relevant to each candidate. Two runs were
performed, each of them made of two Markov chains of 50,000 steps each.
The purpose of the first run was to estimate the level of correlated
noise in each lightcurve and to provide the second run with updated
error bars on the jump parameters. In the second MCMC, the good mixing
and convergence of the Markov chains were assessed using the
Gelman-Rubin statistic criterion \citep{GelmanRubin1992}.

We divided the total lightcurve in chunks of duration of $\sim$24 to 48~hr
and fitted for each of them the smooth photometric variations due to stellar 
variability or instrumental systematic effects with a time-dependent quadratic
polynomial. Baseline model coefficients were determined at each step of the MCMC for each
lightcurve with the singular value decomposition method
\citep{Press1992}. The resulting coefficients were then used to correct the raw photometric lightcurves.
 
For each chunk of data, correlated noise was accounted for following 
\citet{Winn2008,Gillon2010}, to ensure reliable error bars on the fitted parameters. 
For this purpose, we compute a scaling factor based on the standard deviation of
the binned residuals for each lightcurve with different time bins. The error bars are 
then multiplied by this scaling factor. 

The rest of the important inputs for the MCMC is as follows.  
For each quarter, we estimated the degree of photometric dilution by using
the contamination factor\footnote{Contamination values can be found in the
fits files headers.} computed from the KIC crowding matrix \citep{Bryson2010} and
was then applied to the transit photometry.

We assumed a quadratic law for the limb-darkening (LD) and used
$c_1=2u_1+u_2$ and $c_2=u_1-2u_2$ as jump parameters, where $u_1$ and
$u_2$ are the quadratic coefficients. $u_1$ and $u_2$ were drawn from
the theoretical tables of \citet{Claret2011} for the corresponding KIC
$T_{eff}$ and log $g$ values.

The MCMC has the following set of jump parameters: the planet/star
flux ratio, the impact parameter $b$, the transit duration from first
to fourth contact, the time of minimum light $T_0$, the orbital
period, the occultation depth, the two LD combinations $c_1$ and $c_2$
and the two parameters $\sqrt{e}\cos\omega$ and
$\sqrt{e}\sin\omega$. A uniform prior distribution is assumed for all
jump parameters but $c_1$ and $c_2$, for which a normal prior
distribution is used, based on theoretical tables.

\subsection{False positive assessment via secondary eclipses}

The {\it Kepler} giant planet candidates list is not guaranteed against
false positives, although a ranking of preliminary assessment is
provided on MAST. Yet it is the false positive rate contributed by eclipsing binary stars
with larger radii than Jupiter's that would contaminate our
findings. Furthermore, late type M dwarfs could actually produce
planet-to-star area ratio indistinguishable from {\it bona fide} giant
planets, because of their similar radii to Jupiter-like objects
\citep[see, e.g.,][]{Chabrier2009}. Obtaining radial-velocity
measurements at orbital quadrature for more than 100 giant planet candidate objects
is unrealistic for the purpose of the present study, given the faint V magnitude
of the host stars and the number of targets of higher priority in the \textit{Kepler}
follow-up program.

We therefore used our MCMC method to search for a secondary
eclipse whose depth would be indicative of a stellar companion instead of a planet. 
No constraint on the eccentricity was imposed since binaries
with orbital periods of a few days only and eccentricity $e>0.2$ are
not uncommon \citep[see, e.g.,][]{Rucinski2007,Maceroni2009}.

We used the derived occultation depth $\frac{F_P}{F_{\star}}$ to compute the corresponding geometric
albedo $A_g=\frac{F_P}{F_{\star}}\frac{a^2}{R_p^2}$ and the brightness temperature to assess the nature 
of each KOI. We further visually
inspected the individual folded lightcurves, as discontinuities due to
spacecraft roll, change of focus, pointing offsets or safe mode events
could create artifacts in the raw photometry and affect the
detection of shallow features in the lightcurve. We present the
results of this analysis, in terms of false positive rate, in Sect.~3.

\subsection{Stellar parameters}

The photometric calibration of the {\it Kepler} field target stars
presented in the KIC yields stellar radii uncertainties of 35\% RMS
\citep{Brown2011}. Because the stellar radius uncertainty translates
directly to a planet candidate size, the stellar radius
uncertainty is too large for any useful constraint on the behavior
of planetary radii with incident stellar flux and orbital
distance. Hence this motivated us to derive our own stellar radii by
a different method than assuming the KIC stellar radii, which yields smaller
uncertainties (of $\sim$15\% RMS) on the derived stellar radius (see Sect.~3).

The method we use employs the
empirical calibration law presented in \citet{Torres2010}.  The authors
show that accurate stellar masses and radii could be deduced from the
stellar effective temperature $T_{eff}$, surface gravity log $g$ and
metallicity [Fe/H] derived from spectroscopy. For this purpose they
build a calibration law based on a large sample of well characterized
detached binaries. A linear regression algorithm then provides the
stellar mass, in function of the spectroscopic parameters.

\citet{Enoch2010} further suggested to use as input the stellar density $\rho_{\star}$ 
instead of the stellar surface gravity log $g$.
The advantage of this approach is that the stellar density is well constrained
by the transit lightcurve photometry \citep{SeagerMallen2003}, 
yielding better results than using the surface gravity derived
from the spectroscopic analysis.

The empirical calibration implemented in the MCMC is therefore a function of
$T_{eff}$, $\rho_{\star}$ and the stellar metallicity [Fe/H], which is
poorly constrained from the KIC photometry \citep{Brown2011}.  We thus imposed a 0.3 dex
uncertainty on the stellar metallicity.  At each step of the MCMC,
$\rho_{\star}$ (deduced from the jump parameters), $T_{eff}$ and
[Fe/H] (drawn from the normal distribution based on the KIC value with
the error bars quoted above) are used as input to the calibration
law. The physical parameters of the system are then deduced
using the resulting stellar mass. The intrinsic uncertainty of the
parameters of the calibration relationship is accounted for by
randomly drawing the parameters values from the normal distribution
presented in \citet{Torres2010} at each iteration of the MCMC. 
The remainder of the uncertainty on the stellar radius is then mostly dominated 
by the error on the KIC $T_{eff}$ and on the intrisic scatter of the empirical 
relationship.  

This method makes the derivation of the stellar mass possible 
at each step of the MCMC without the need of performing a
separate analysis based on stellar evolution models.

\section{Results}

The main result of this study is that giant planet candidate radii are independent
of stellar incident flux below an incident flux of about $2\times
10^8$ erg s$^{-1}$cm$^{-2}$ (Figure~1). Although the giant planet radius trend was hinted
  with published giant planets alone \citep[][see also Fig.~1]{MillerFortney2011} and theoretically
  expected \citep[e.g.,][]{Fortney2007}, inclusion of 
  these new Kepler giant planet candidates radii yields a robust trend.

The objects supporting this result are the {\it Kepler} giant planet
candidates that have no or shallow secondary eclipses consistent with
their equilibrium temperature at 2$\sigma$ level or less.  For
comparison we include transiting planets not discovered
by {\it Kepler} which overlap perfectly in the $R_{planet}$ vs. incident flux plane, 
but mainly populate the high
incident flux regime.  The {\it Kepler} Q0-Q2 coverage not only
enables almost a doubling of the transiting giant planet candidates but
also extends the coverage out to lower incident fluxes as compared to
currently known transiting planets.

Complicating the result is the fact that {\it Kepler} planet
candidates are not vetted as actual planets. We have used our MCMC
analysis to assess false positives via secondary eclipse
detection. Indeed, 16 planet candidates show strong evidence for
deep secondary eclipses suggesting a 4$\sigma$ discrepancy or more with their
estimated equilibrium temperature.
Such objects are discarded from the study. Finally, 22 planet candidates yield
a secondary eclipse signature whose origin cannot be
secured, the inferred brightness temperature being consistent with either a planetary or stellar companion. We still choose to include those candidates in Figure~1, with distinct symbols. Additional data will help in tightening the nature of those objects.

We notice that eclipsing binaries with grazing transits combined to non-zero orbital 
eccentricity would not yield any secondary eclipse and would therefore be
wrongly identified as planets in our study. Any such contamination 
should be uniform with the range of incident fluxes explored in this study and would
not affect the main finding of a trend in giant planet radii.

For most of our planet candidates, there is no information about
the orbital eccentricity. Instead of unrealistically assuming circular orbits, we
assigned each candidate an eccentricity value drawn from the distribution
presented in \citet{Wang2011}, as well as a random value for the argument of periastron.
This approach is reasonable since no significant trend seems to exist between
eccentricity and orbital period for \textit{Kepler} candidates \citep{Moorhead2011}.

To gauge the impact of orbital eccentricity on our results,
we performed a new MCMC analysis by imposing priors on
$\sqrt{e}\sin\omega$ and $\sqrt{e}\cos\omega$ (see Sect.~2.2),
based on the values drawn for $e$ and $\omega$ in the previous step.
We then computed the orbit-averaged incident flux for each candidate
and found an excellent agreement (4\% in average) with the fluxes 
obtained for the circular case.
The reported trend in the R$_p$ vs. incident flux plane is therefore robust
to the planetary candidates' orbital eccentricity. We show our results (assuming
the eccentricity distribution described above) on Fig.~1.

In summary, out of the 115 KOI, 70 of them exhibit no or shallow secondary eclipses consistent
with their equilibrium temperature. Those objects are therefore
considered as of planetary origin, whereas 16 are identified as stellar companions and 22 are of  
ambiguous classification. We list the KOIs and their identification in Table~1. 

In the course of the analysis, 7 KOI got their radius revised to 
less than 6 $R_{\oplus}$. Those KOI were thus discarded from our study, as no longer part of the giant planet candidate sample.

\begin{deluxetable}{cccc}
\tabletypesize{\scriptsize}
\tablecomments{List of KOI used in this study. KOI meeting our criteria (see Sect.~3) are shown in the ``planetary" column while the probable stellar companions are shown in the ``False positives" column. The intermediate class is shown under ``ambiguous".}
\tablenum{1}
\tablehead{\colhead{Planetary} & \colhead{Planetary (continued)} & \colhead{Ambiguous} & \colhead{False Positives} }
\startdata
\tableline
1.01 & 398.01 & 12.01 & 194.01\\
2.01 & 410.01 & 187.01 & 197.01\\
10.01 & 417.01 & 189.01 & 208.01\\
17.01 & 418.01 & 458.01 & 552.01\\
18.01 & 421.01 & 617.01 & 609.01\\
20.01 & 423.01 & 728.01 & 743.01\\
22.01 & 425.01 & 763.01 & 745.01\\
94.01 & 625.01 & 767.01 & 779.01\\
97.01 & 674.01 & 772.01 & 876.01\\
98.01 & 686.01 & 823.01 & 895.01\\
100.01 & 698.01 & 840.01 & 1003.01\\
127.01 & 760.01 & 855.01 & 1152.01\\
128.01 & 801.01 & 856.01 & 1177.01\\
135.01 & 805.01 & 918.01 & 1540.01\\
138.01 & 806.01 & 929.01 & 1541.01\\
183.01 & 806.02 & 960.01 & 1543.01\\
186.01 & 809.01 & 961.02 & \\
188.01 & 815.01 & 961.03 & \\
190.01 & 824.01 & 1020.01 & \\
191.01 & 846.01 & 1285.01 & \\
192.01 & 850.01 & 1299.01 & \\
193.01 & 858.01 & 1385.01 & \\
195.01 & 871.01 &  & \\
196.01 & 882.01 &  & \\
199.01 & 883.01 &  & \\
202.01 & 889.01 &  & \\
203.01 & 897.01 &  & \\
205.01 & 908.01 &  & \\
214.01 & 913.01 &  & \\
217.01 & 931.01 &  & \\
254.01 & 1089.01 &  & \\
351.01 & 1176.01 &  & \\
366.01 & 1227.01 &  & \\
368.01 & 1391.01 &  & \\
372.01 & 1486.01 &  & \\
\enddata
\end{deluxetable}

\begin{figure}
\epsscale{1.25}
\plotone{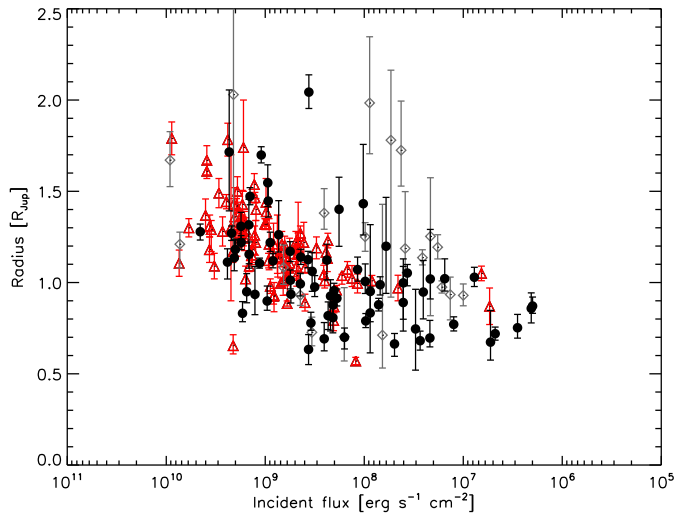}
\caption{Planetary radii as a function of incident flux. Black filled circles are KOI ranked as
planetary candidates in the frame of this work while gray diamonds represent KOI whose
origin is ambiguous (see Sect.~3). 
Transiting giant planets previously published, and mostly from ground-based surveys, are shown as red triangles. The relevant parameters $R_p$, $R_s$, $T_{eff}$ and $a$ have been drawn from http://www.inscience.ch/transits on August 29, 2011.  \label{fig1}}
\end{figure}

\section{Discussion}

\subsection{Giant planet radii inflation mechanisms}

KOI classified as planetary candidates in the
frame of this study yield planetary radii vs. incident flux in good
agreement with published transiting planet data, as shown on Fig.~1.  Remarkably, this set of
KOI results in constant radii of $\sim$ 0.87$\pm$0.12$R_J$
\footnote{Jupiter radius itself is 0.977 $R_J$ when using its mean radius of 69,894 km
instead of its equatorial radius of 71,492 km \citep[see Table~1 of][]{Guillot2005}}, 
similar to Jupiter, below an incident flux of $\sim$ $2\times 10^8 $erg s$^{-1}$cm$^{-2}$.  
We report no inflated giant planet radii below this threshold.

Several explanations have been invoked to bring or maintain heat in the planetary
interiors, necessary to explain radii anomalies through a larger equilibrium radius.
An exhaustive description can be found in, e.g., \citet{FortneyNettelmann2010, Baraffe2010} but none
seems to reproduce all planets with inflated radii. 

Tidal energy dissipation in the giant
planet interior is expected to counteract the contraction \citep{Bodenheimer2001}. 
\citet{Bodenheimer2003} proposed that an additional
companion in the system could pump the planet eccentricity that would be dissipated
through tides. Significant follow-up on this possibility emerged in the last years. 
\citet{Levrard2009} have for instance shown that most known transiting
planets were spiraling toward their star due to tidal dissipation. 
\citet{Miller2009} modeled coupled thermal evolution and tidal effects on 
giant planets and were able to reproduce the radius anomalies for 35 out of 
45 planets part of their sample, assuming {\it ad hoc} initial conditions.

Electrical current
generated through the interaction of ionized particles with the
planetary magnetic field causes a dissipation of energy in the
planetary interior \citep{BatyginStevenson2010}. 
\citet{Laughlin2011} find support for this
hypothesis from the set of transiting planets known in 2010 but also
state that other processes should be contributing to account for the
observed anomalies, such as the effects of heavy elements abundances \citep{Batygin2011} 
or the internal heating induced by tidal circularization for eccentric planets.

Layered convection should occur 
in atmospheres characterized by molecular weight gradients \citep{ChabrierBaraffe2007}.
This would decrease
the loss of heat and slow down the contraction in the planetary interiors. This
mechanism is independent of the incident stellar flux.

Our results suggest that the combinations of mechanisms responsible for the giant
planets inflated radii are correlated to the strength of the stellar incident flux. 

For the most close-in planetary candidates, tidal effects could contribute significantly and add to the sole incidence of stellar irradiation.

As part of this study, we find no support for a process that would be solely based on layered 
convection as no cold inflated planets are reported.

There is observed scatter in the plateau of giant planets radii below $\sim$ $2\times 10^8$ erg s$^{-1}$cm$^{-2}$. We speculate this scatter could be due to the effects of metallicity. Planets enriched with heavy elements yield
a more compact structure, thus a smaller radius, like HD\,149026\,b \citep{Sato2005}.

\subsection{False positives rate}

A byproduct of our MCMC analysis is also an important result: an
estimated false positive rate for the {\it Kepler} giant planet
candidate listing. According to our analysis, 14\% of the {\it Kepler} giant planet candidates
studied in this sample are eclipsing binaries or background eclipsing
binaries. This result is based on identification of a large discrepancy between 
the candidate's equilibrium temperature and its measured brightness temperature 
in the {\it Kepler} bandpass. We further note that the 14\% false positive rate number
might even be higher in the case of grazing events or eccentric
orbits, for which the detection of a secondary eclipse is not possible. We
found no false positives among the 10 multi-planetary systems included
in our sample.

Additional {\it Kepler} data will allow an extension of the present study
to planets with longer orbits than the candidates presented in this work.
The next quarters of {\it Kepler} photometry
will also help in tightening the exact fraction of false positives among
giant planet candidates by improving the characterization of secondary eclipses.

In summary, this work presents one of the first results from the emerging science of exoplanet statistics enabled by {\it Kepler}'s exquisite photometry and large pool of planet candidates. With future {\it Kepler} data, we expect many other planet population trends to be identified and to weigh in or solve key exoplanet questions.



\acknowledgments 
We are grateful to Jonathan Fortney, Jack Lissauer, Eric Ford and J\'er\'emy Leconte
 for insightful comments that improved this manuscript. We warmly thank Micha\"el Gillon for
sharing his expertise on MCMC methods. We thank the \textit{Kepler} Giant Planet
Working Group for useful discussions and especially Jason Rowe and
Jon Jenkins for their invaluable inputs regarding \textit{Kepler} photometry.
We thank the anonymous referee for an helpful review that improved this manuscript.
Funding for the Kepler Mission is provided by the National Aeronautics and Space Administration (NASA) Science Mission Directorate. This work was in part funded by the Kepler Participating Science Program grant NNX08BA51G. 
{\it Facility:} \facility{Kepler.}


\clearpage

\end{document}